\documentstyle[12pt,graphicx]{article}
\begin{document}
\title{Kaluza-Klein Magnetic Monopole in Five-Dimensional Global Monopole
Spacetime}
\author{A. L. Cavalcanti de Oliveira \thanks{E-mail: alo@fisica.ufpb.br} \ 
and E. R. Bezerra de Mello \thanks{E-mail: emello@fisica.ufpb.br}
\\
Departamento de F\'{\i}sica-CCEN\\
Universidade Federal da Para\'{\i}ba\\
58.059-970, J. Pessoa, PB\\
C. Postal 5.008\\
Brazil}
\maketitle
\begin{abstract}
In this paper we present a solution for Kaluza-Klein magnetic monopole in a
five-dimensional global monopole spacetime. This new solution is a generalization
of the previous ones obtained by D. Gross and M. Perry (Nucl. Phys. B {\bf 226}, 29 
(1983)) containing a magnetic monopole in a Ricci-flat formalism, and by A. Banerjee,
S. Charttejee and A. See (Class. Quantum Grav. {\bf 13}, 3141 (1996)) for a global
monopole in a five-dimensional spacetime, setting specific integration
constant equal to zero. Also we analyse the classical motion of a massive charged
test particle on this manifold and present the equation for classical trajectory 
obeyed by this particle. 
\end{abstract}

\newpage
\renewcommand{\thesection}{\arabic{section}.}
\section{Introduction}

One of the oldest and elegant formalism to unify the gravitation with the gauge
theory was proposed by Th. Kaluza \cite{Kaluza} many years ago. Kaluza's conjecture
was that the degrees of freedom associated with the gauge field could be 
accommodated as the new components of the metric tensor in a higher than four 
dimensional manifold. Specifically, considering the Abelian gauge theory, just one 
extra  dimension would be enough. This extra dimension is compactified on a circle
of so small radius that would not be observable at low-energy scale, i.e., smaller
than the Planck one. This theory was analyzed by several authors including O. Klein,
who clarified many aspects of the structure of the manifold \cite{Klein}.

Five-dimensional Einstein action exhibits, as the low-energy effective theory, the 
four-dimensional gravity theory coupled with Maxwell one, where all the physical
fields do not depend on the fifth coordinate. The generalization of the theory to 
include non-Abelian gauge fields requires the addition of more than one extra 
dimensions. 

Also one of most important works about Abelian gauge theories was due to the P. M. 
Dirac many years ago, who proposed a new solution to the Maxwell equations. His new
solution for the vector potential corresponds to a point-like magnetic 
monopole with a singularity string running from the particle's position to infinity
\cite{Dirac}. The most elegant formalism to describe the Abelian point-like magnetic
monopole has been developed by Wu and Yang \cite{Wu-Yang}. In their formalism the 
vector potential is described by a singularity free expression. In order to provide 
this formalism, Wu and Yang defined the vector potential $A_\mu$ in two overlapping 
regions, $R_a$ and $R_b$, which cover the whole space. 

In their beautiful papers, Gross and Perry \cite{G-P}, and Sorkin \cite{Sorkin}, 
independently, presented a soliton-like solution of the five-dimensional 
Kaluza-Klein theory corresponding to a magnetic monopole. As the Dirac solution, 
their solutions describes a gauge-dependent string singularity line, if the fifth 
coordinate is conveniently compactified. Moreover, its magnetic charge has one unit 
of Dirac charge: $g=1/2e$ in units $\hbar=c=1$. These solutions are generalizations 
of the self-dual Euclidean Taub-NUT solution \cite{T-NUT}. Also, Gegenberg and 
Kunstatter in \cite{G-K} found another magnetic monopole solution for 
five-dimensional Kaluza-Klein theory. Their solutions were obtained by applying 
the static and Ricci-flat requirement on the field equation.

Global monopole is a solution predicted in Grand Unified Theories. It is 
formed due to a phase transition of a system composed by self-coupling iso-scalar
field only. The matter field plays the role of an order parameter which outside 
the monopole's core, acquires a non-vanishing value. The simplest theoretical model 
which gives rise to global monopole has been proposed by Barriola and Vilenkin 
\cite{B-V}. This model is composed by triplet Goldstone field $\phi^a$. The original 
global $O(3)$ symmetry of the physical system is spontaneously broken down to $U(1)$.
In four-dimensional spacetime this Lagrangian density reads:
\begin{equation}
\label{GM}
{\cal{L}}=-\frac12g^{\mu\nu}\partial_\mu\phi^a\partial_\nu\phi^a-\frac14\lambda
\left(\phi^a\phi^a-\eta^2\right)^2  
\end{equation} 
with $a=1, \ 2, \ 3$ and $\eta$ being the scale energy where the symmetry is broken. 
The field configuration which describes a monopole is
\begin{equation}
\label{phi}
\phi^a(x)=\eta f(r)\hat{x^a} \ ,
\end{equation}
where $\hat{x^a}\hat{x^a}=1$. Coupling this matter field with the Einstein equation, 
a spherically symmetric regular metric tensor solution is obtained. Barriola and
Vilenkin also shown that for points outside the global monopole's core the geometry 
of the manifold can be approximately given by the line element
\begin{equation}
\label{gm} 
ds^2=-dt^2+\frac{dr^2}{\alpha^2}+r^2(d\theta^2+\sin^2\theta d\phi^2)  
\end{equation}
with $\alpha^2=1-8\pi G\eta^2$. This line element represents a three-geometry with 
a solid angle deficit.

The analysis of the above system in the context of a five-dimensional Einstein 
equation has been developed by Banerjee {\it at al} in \cite{BCS}. There a family of 
solutions has been found in the region outside the global monopole, where the
Goldstone boson can be approximated by $\phi^a(x)=\eta\hat{x^a}$. They pointed out 
that, differently from the Barriola and Vilenkin solution, the criteria of 
uniqueness is lost in five dimensions, and for specific choice of parameters their 
solution is Schwarzschild-type one.

Here we shall continue the analysis developed by Banerjee {\it at al} admitting 
the presence of a magnetic monopole. As we shall see, the presence of the global 
monopole system gives rise to a non Ricci-flat solution. So, in order to make this 
analysis possible, we shall consider the case where both defects are at the same 
position chosen as the origin of the reference system. In this way our solution 
contains the solutions found by Gross and Perry and Banerjee {\it et al} as special 
cases. 

The analysis of a system that presents a regular composite topological object, 
which takes into account the presence of a self-gravitating 't Hooft-Polyakov 
magnetic monopole in a global monopole spacetime, has been developed recently by one 
of us in \cite{Spi}. There it was found that at large distance the structure of the 
manifold corresponds to a Reissner-Nordst\"om spacetime with a solid angle deficit 
factor. Here we shall present also a composite monopole solution in the context of
a five-dimensional Einstein equation. Our solution contains an Abelian magnetic
monopole in the presence of a five-dimensional global monopole spacetime.

This paper is organized as follows: In Sec. $2$ we briefly review the results found
by Gross and Perry and Banerjee {\it et al}, we introduce the mathematical
formalism needed to develop our analysis and present the complete system that we want
to study. Also we present our solutions obtained from the Einstein equations in five 
dimensions taking into consideration the presence of the Abelian magnetic monopole. 
In Sec. $3$ we study the classical relativistic motion of a test charged particle 
on this manifold and present the equation for the trajectory obeyed by it.  Finally 
we present in Sec. $4$ our conclusions and most relevant remarks. 

\section{Composite Monopole}

As we have already said, in this section we analyse the physical system
given by (\ref{GM}) in the context of a five-dimensional Einstein equation. In
order to take into account the presence of an Abelian magnetic monopole we have
to admit non-diagonal components to the metric tensor. Following the prescription
adopted by \cite{G-P}, we shall assume static spherically symmetric structure for 
the four-dimensional spacetime components of the metric tensor. Using the
coordinates ${\hat{x}}^A=(t,\ r, \ \theta, \ \phi, \ \Psi)$, with the index $A$ 
running from $0$ to $4$ the line element reads: 
\begin{eqnarray}
\label{metric}
d{\hat{s}}^2&=&{\hat{g}}_{AB}d{\hat{x}}^Ad{\hat{x}}^B=-B(r)dt^2+A(r)dr^2+C(r)r^2
\left(d\theta^2+\sin^2\theta d\phi^2\right)\nonumber\\
&+&D(r)(d\Psi+f(\theta)d\phi)^2 \ .
\end{eqnarray}
As we can see the structure of the above expression can represent two special cases
of solutions of the five-dimensional Einstein equation as shown in the following:

\subsection{Gross and Perry Solution}

Admitting the absence of the global monopole, i.e., taking $\eta=0$ in 
(\ref{GM}), the Ricci-flat, ${\hat{R}}_{AB}=0$, solution found by Gross and Perry
is:
\begin{equation}
A(r)=C(r)=\frac1{D(r)}=1+\frac{4m}r \ , \ \ B(r)=1 \ , \ \ f(\theta)=
4m(1-\cos\theta) \ .
\end{equation}
In order to have singularity-free solution, the parameter $m$ is chosen to be $R/8$,
being $R$ the radius of the circle of the fifth dimension. Under this condition the
magnetic charge of the monopole, $g$, is related with $m$ parameter by 
$m=g\sqrt{\pi G}$ \cite{G-P}. Here $G$ denotes Newton's gravitational constant.

Moreover, Gross and Perry called attention to the fact that although the 
soliton-solution described by the above expressions presents no gravitational
mass, it does possess an inertial mass of order of the Planck one.

\subsection{Banerjee {\it et al} Solution}

In the absence of the magnetic monopole, $f(\theta)$ is taken to be zero. In this 
case (\ref{metric}) becomes diagonal. The solutions found by Banerjee {\it et al} 
to the Einstein equations in the presence of the system described by (\ref{GM})
were obtained in the region outside the global monopole, 
where, approximately, $f(r)\simeq 1$. Doing this, they were able to find a family 
of solutions:
\begin{eqnarray}
B(r)=\left(\alpha^2-\frac{2G{\cal{M}}}r\right)^a \ , \ \ A(r)=\left(\alpha^2-
\frac{2G{\cal{M}}}r\right)^{-(a+b)}\nonumber\\
C(r)=\left(\alpha^2-\frac{2G{\cal{M}}}r\right)^{(1-a-b)} \ , \ \ D(r)=\left(\alpha^2
-\frac{2G{\cal{M}}}r\right)^b \ ,
\end{eqnarray}
where ${\cal{M}}$ is a constant of integration, and $a$ and $b$ are two 
dimensionless parameters which obey the consistency condition $a^2+ab+b^2=1$. So, 
differently from the solution found by Barriola and Vilenkin in a four-dimensional 
spacetime, in this formalism the uniqueness criteria to the solution is lost. For 
the particular choice $a=1$ and $b=0$, the above solution can be understood as a 
five-dimensional extension of the Barriola and Vilenkin one.

\subsection{The Model}

After the above review about these two models we shall present our model. We shall
consider the physical system which presents an Abelian magnetic monopole in
a five-dimensional global monopole manifold. We shall see that due to the presence 
of latter system, the Ricci-flat condition is not longer fulfilled. In order to
proceed with our investigation, we first present the non-vanishing components 
of the energy-momentum tensor associated with the global monopole Lagrangian density 
(\ref{GM}), considered as an external source:
\begin{eqnarray}
T_{00}&=&-\eta^2B(r)\left[\frac{f'^2(r)}{2A(r)}+\frac{f^2(r)}{r^2C(r)}+
\frac{\lambda\eta^2}4(f^2(r)-1)^2 \right] \ , \nonumber\\
T_{11}&=&-\eta^2\left[\frac{f'^2(r)}2-A(r)\frac{f^2(r)}{r^2C(r)}
-\frac{\lambda\eta^2}4A(r)(f^2(r)-1)^2 \right] \ , \nonumber\\
T_{22}&=&\eta^2r^2C(r)\left[\frac{f'^2(r)}{2A(r)}+\frac{\lambda\eta^2}4(f^2(r)-1)^2
\right] \ , \nonumber\\
T_{33}&=&\eta^2r^2C(r)\sin^2\theta\left[\frac{f'^2(r)}{2A(r)}+\frac{\lambda\eta^2}4
(f^2(r)-1)^2\right]\nonumber\\
&+&\eta^2D(r)f^2(\theta)\left[\frac{f'^2(r)}{2A(r)}+\frac{f^2(r)}
{r^2C(r)}+\frac{\lambda\eta^2}4(f^2(r)-1)^2 \right] \ , \nonumber\\
T_{34}&=&T_{43}=\eta^2D(r)f(\theta)\left[\frac{f'^2(r)}{2A(r)}+\frac{f^2(r)}{r^2C(r)}
+\frac{\lambda\eta^2}4(f^2(r)-1)^2\right] \ , \nonumber\\
T_{44}&=&\eta^2D(r)\left[\frac{f'^2(r)}{2A(r)}+\frac{f^2(r)}{r^2C(r)}+
\frac{\lambda\eta^2}4(f^2(r)-1)^2\right] \ .
\end{eqnarray}

The equation for the Higgs field in the metric (\ref{metric}) gives rise to the
following radial differential equation:
\begin{eqnarray}
\frac1{A(r)}f''(r)+\left[\frac2{A(r)r}+\frac1{2B(r)C^2(r)D(r)}\left(\frac{B(r)C^2(r)
D(r)}{A(r)}\right)'\right]f'(r)\nonumber\\
-\frac2{C(r)r^2}f(r)-\lambda\eta^2f(r)(f^2(r)-1)=0 \ .
\end{eqnarray}

In four-dimensional flat spacetime the similar differential equation has no 
analytical solution; however it is shown \cite{B-V} that for radial distance $r$
larger than the monopole's core, $\delta \simeq \lambda^{-1/2}\eta^{-1}$, $f(r)
\simeq 1$, and for $r\to 0$, $f(0)=0.$ We observe that in this present context the
same boundary conditions can be applied to the function $f(r)$. So in the follows
we shall analyse the complete system in the region outside the global monopole's
core. In this way the components for the energy-momentum tensor become much 
simpler. They read:
\begin{eqnarray}
T_{00}=-\eta^2\frac{B(r)}{r^2C(r)} \ , \ \ T_{11}=\eta^2\frac{A(r)}{r^2C(r)} \ ,
\ \ T_{22}=0 \ , \nonumber\\
T_{33}=\eta^2\frac{f^2(\theta)D(r)}{r^2C(r)} \ , \ \
T_{34}=\eta^2\frac{D(r)f(\theta)}{r^2C(r)} \ , \ \ T_{44}=\eta^2\frac{D(r)}{r^2C(r)} 
\ .
\end{eqnarray}

From the five-dimensional Einstein equation,
\begin{equation}
\label{EE}
{\hat{R}}_{AB}=8\pi G_K\left({\hat{T}}_{AB}-\frac{{\hat{g}}_{AB}}3{\hat{T}}\right) 
\ ,
\end{equation}
where $G_K$ is the five-dimensional gravitational constant, we find that the only 
non-vanishing components of the Ricci tensor are:
\begin{equation}
\label{Ricci} 
R_{22}=\alpha^2-1 \ , \ \ R_{33}=(\alpha^2-1)\sin^2\theta \ .
\end{equation}
Here $\alpha^2=1-8\pi G_K\eta^2$. However it is possible to relate the 
five-dimensional gravitational coupling constant with the Newton's one $G$ by
\begin{equation}
G_K=2\pi RG \ .
\end{equation}
Defining the energy scale $\eta$ in the five-dimensional spacetime as the ordinary 
one in four-dimensions divided by the $\sqrt{2\pi R}$, we re-obtain for the parameter
$\alpha$ the same expression as given before to the Barriola and Vilenkin model.

Now, to complete this analysis, we have to find solutions for the functions $B(r)$, 
$A(r)$, $C(r)$, $D(r)$ and  $f(\theta)$ compatible with the above results. Because 
we want that our expression would reproduce a four dimensional generalization of 
the self-dual Euclidean Taub-NUT solution in presence of a global monopole, we must 
have $B(r)=1$. Moreover, because it must approach asymptotically to the 
five-dimensional extension of the Barriola and Vilenkin solution found in Ref. 
\cite{BCS}, $\alpha^2$ should be a multiplicative factor. Although the existence of
the magnetic monopole depends on the topology of the spacetime, in this case, 
supported by previous analysis about composite topological defect \cite{Spi}, we 
can infer that the presence of the global monopole does not modify the configuration 
of the magnetic monopole. Taking all these informations in consideration we find:
\begin{eqnarray}
\label{comp}
A(r)&=&(\alpha^2)^{-\frac{1+a}2}\left(1+\frac{4m}{\alpha r}\right) \ , \nonumber\\
C(r)&=&(\alpha^2)^{\frac{1-a}2}\left(1+\frac{4m}{\alpha r}\right) \ , \nonumber\\
D(r)&=&(\alpha^2)^{\frac{1-a}2}\left(1+\frac{4m}{\alpha r}\right)^{-1} \ , 
\nonumber\\
f(\theta)&=&4m(1-\cos\theta) 
\end{eqnarray}
for any value of the parameter $a$ \footnote{The complete set of field equations
derived from (\ref{EE}) presents very longs expressions, even considering specific 
ansatz to the unknown functions. This is the reason why we decided do not include 
this set of differential equation in our paper.}. At this point we could think 
that our solutions, as the Banerjee {\it et al} ones, represent a family of 
independent solutions. However, this is not true: by a global scale transformation 
on the metric tensor, ${\hat{g}}_{AB} \to \alpha^{a-1}{\hat{g}}_{AB}$ and redefining 
the time coordinate
appropriately, the following line element is obtained:
\begin{eqnarray}
\label{metric1}
d{\hat{s}}^2&=&-dt^2+V(r)\left(\frac{dr^2}{\alpha^2}+r^2(d\theta^2+\sin^2\theta 
d\phi^2)\right)\nonumber\\
&+&V(r)^{-1}(d\Psi+4m(1-\cos\theta)d\phi)^2 
\end{eqnarray}
with
\begin{equation}
V(r)=1+\frac{4m}{\alpha r} \ .
\end{equation}

This solution presents the most relevant properties associated with the global and 
magnetic monopoles: $i)$ it can be considered as a five-dimensional extension of the 
Barriola and Vilenkin solution, in the sense that the space section asymptotically 
presents a solid angle $\Omega=4\pi\alpha^2$, consequently smaller than the ordinary 
one, and $ii)$ also presents an Abelian magnetic monopole.

As it was pointed out by Gross and Perry in Ref. \cite{G-P}, the gauge field 
associated with the magnetic monopole 
\begin{equation}
A_\phi=4m(1-\cos\theta) \ ,
\end{equation} 
presents a singularity at $\theta=\pi$. However, this singularity is gauge dependent
if the period of the compactified coordinate $\Psi$ is equal to $16\pi m$. This is 
the geometric description of the Dirac quantization. Adopting this period for the 
extra coordinate, it is possible to provide the Wu and Yang formalism to describe 
the four-vector potential, $A_\mu$, associated with the Abelian magnetic monopole 
without line of singularity. In order to do that it is necessary to construct two 
overlapping regions, $R_a$ and $R_b$, which cover the whole space section of the 
manifold. Using spherical coordinate system, with the monopole at origin the only 
non-vanishing components for the vector potential are
\begin{eqnarray}
\label{A}
(A_\phi)_a&=&4m(1-\cos\theta) \ ,\  R_a: \ 0 \leq \theta < \frac12\pi+\delta \ ,
\nonumber\\
(A_\phi)_b&=&-4m(1+\cos\theta) \ ,\ R_b: \ \frac12\pi-\delta< \theta \leq \pi \ , 
\end{eqnarray}
with $0<\delta <\pi/2$. In the overlapping region, $R_{ab}$, the non-vanishing 
components are related by a gauge transformation. Using the appropriate normalization
factor \cite{G-P}, one can rewrite the above vector potential in terms of the 
physical one, $A^{ph}_\phi$:
\begin{equation}
\sqrt{16\pi G}(A^{ph}_\phi)_a=\sqrt{16\pi G}\left[(A^{ph}_\phi)_b+
\frac ieS\partial_\phi S^{-1}\right] \ ,
\end{equation}
where $S=e^{2i\omega\phi}$, $\omega=-eg=-n/2$ in units $\hbar=c=1$ and $g$ being 
the monopole strength. In terms of non-physical vector potential this gauge 
transformation corresponds to subtract the quantity $8m$, which compensates the 
changing in the fifth coordinate $\Psi'=\Psi+8m\phi$. Also we must say that the 
same Ricci tensors (\ref{Ricci}) are obtained for both expressions of the 
four-vector potential.

Before to finish this section three important remarks about the solution should 
be made: $i)$ The radial function $V(r)$ in the line element 
(\ref{metric1}) differs from the similar one found by Gross and Perry by $\alpha$ 
factor multiplying the radial coordinate in the denominator. The obtained
magnetic field is
\begin{equation}
{\vec{B}}={\vec{\nabla}}\times{\vec{A}}=\frac1{V(r)}\frac{4m}{r^2}{\hat{r}}
=-\frac1{\sqrt{V}}\vec{\nabla}V(r) \ ,
\end{equation}
which asymptotically gives rise to the usual Dirac magnetic monopole. Moreover, we 
can see by calculating the total magnetic flux on a spherical surface concentric 
with the monopole, that $\Phi_B=4m(4\pi)$. $ii)$ Changing the sign of $m$ in $V(r)$, 
we obtain another solution of the field equations. $iii)$ Finally, we want 
to emphasize that the solutions found for the components of the metric tensor are 
valid only in the region outside the global monopole.

\section{Analysis of the Motion of a Charged Particle in the Manifold}

As we have already said, the line element (\ref{metric1}) is the five-dimensional
extension of the Barriola and Vilenkin solution in the presence of an Abelian 
magnetic monopole. The classical motion of a test massive particle in this manifold 
can be analysed by a Lagrangian obtained by differentiating this quantity with 
respect to some affine parameter $\xi$:
\begin{eqnarray}
\label{Lagrangian}
L&=&-\dot{t}^2+\left(1+\frac{4m}{\alpha r}\right)\left(\frac{\dot{r}^2}{\alpha^2}+
r^2(\dot{\theta}^2+\sin^2\theta\dot{\phi}^2)\right)\nonumber\\
&+&\left(1+\frac{4m}{\alpha r}\right)^{-1}(\dot{\Psi}+4m(1-\cos\theta)
\dot{\phi})^2 \ .
\end{eqnarray}

Because the above Lagrangian does not depend explicitly on the coordinates $t$, 
$\Psi$ and $\phi$, three constants of motion can be promptly identified:
\begin{equation}
\dot{t}=a \ ,
\end{equation}
\begin{equation}
\label{k}
V^{-1}(r)\left[\dot{\Psi}+4m(1-\cos\theta)\dot\phi\right]=\kappa \ ,
\end{equation}
and
\begin{equation}
\label{h}
V(r)r^2\sin^2\theta\dot\phi+V^{-1}(r)\left[\dot{\Psi}+4m(1-\cos\theta)\dot\phi\right]
4m(1-\cos\theta)=h \ .
\end{equation}
The equation (\ref{h}) can be written in a simpler form if we use the 
definition of the constant $\kappa$ given in (\ref{k}). Adopting the notation 
given in the paper by Gross and Perry, this constant is the ratio of the charge of 
the test particle to its mass: $\kappa=q/M$. Including in (\ref{h}) the definition  
for the physical magnetic charge and recognizing $q\sqrt{16\pi G}$ as the physical 
charge of the particle, we can identify the $z-$component of the conserved total 
angular momentum associated with a charged particle in this manifold as:
\begin{equation}
J_z=\frac M2h=V(r)Mr^2\sin^2\theta\dot{\phi}+\omega\cos\theta \ ,
\label{Jz}
\end{equation}
where $\omega=-ge$. (In the deduction of the above expression we discarded the 
constant $4m\kappa$ in (\ref{h}).) Moreover the classical equation of motions to the 
polar and radial variables can be obtained, respectively, by the Euler-Lagrange 
formalism and by imposing that the Lagrangian above is a constant $\epsilon$. This
constant can be $0$, $1$ and $-1$, respectively, if the geodesic associated with the 
motion of the particle is null, for massless particle, spacelike and timelike. 
Finally these equations are:
\begin{equation}
\label{theta} 
\ddot{\theta}+\left[\frac2r+\frac{V'(r)}{V(r)}\right]\dot{r}\dot{\theta}
-\sin\theta\cos\theta\dot{\phi}^2-\frac{4m\kappa\sin\theta}{V(r)r^2}\dot{\phi}=0 \ ,
\end{equation}
and
\begin{equation}
\label{r} 
\frac{V(r)}{\alpha^2}\dot{r}^2+V(r)r^2\dot{\theta}^2+V(r)r^2\sin^2\theta
\dot{\phi}^2+\kappa^2V(r)=\epsilon+a^2 \ .
\end{equation}

Schwinger {\it et al} \cite{Schwinger} shown many years ago that the conserved 
total angular momentum, associated with an electric charged particle in the 
presence of a magnetic monopole, is given by
\begin{equation}
\vec{J}=\vec{l}+\omega\hat{r} \ ,
\end{equation} 
with $\vec{l}$ being the ordinary orbital angular momentum. \footnote{In fact it
was H. Poincar\'e \cite{HP} who first investigate the classical motion of an
electron in the presence of a magnetic pole. J. Schwinger and collaborators 
generalized this analysis to two dyons.} Because $\vec{J}\cdot\hat{r}=\omega$, the 
motion of the particle is confined to a cone of half-polar angle $\theta_0$ given by:
\begin{equation}
\cot\theta_0=\frac{|\omega|}{l} \ .
\end{equation}
This means that a particular choice of coordinate system has been adopted, and in 
this system the direction of the $z-$axis is parallel to the vector $\vec{J}$. The
ordinary angular momentum vector $\vec{l}$ is given in terms of the unity vector
$\hat{\theta}$ only: $\vec{l}=-l\hat{\theta}$. In the present case that we are 
analysing we can observe that $\theta=const$ is solution of (\ref{theta}) 
providing
\begin{equation}
\dot{\phi}=-\frac{4mq}{MV(r)r^2\cos\theta}=\frac{\omega}{MV(r)r^2\cos\theta} \ .
\label{lz}
\end{equation}
This result is compatible with what we expected in the sense that the motion of the
particle here is also constrained to a cone. (In particular for $r\to\infty$, 
$V(r)\to 1$, so the above expression reproduces the angular velocity associated with 
a charged particle in a flat spacetime in the presence of a magnetic monopole.)
Moreover, by making a specific choice for the coordinate system we can infer, 
from (\ref{lz}), that the first term on the right hand side of (\ref{Jz}) 
corresponds to the conserved $z-$component of the ordinary angular momentum in this 
manifold and consequently
\begin{equation}
\label{l}
\dot{\phi}=\frac l{MV(r)r^2\sin\theta} \ .
\end{equation}

Finally the equation of motion relating the radial coordinate with the azimuthal 
angle can be obtained combining (\ref{r}) and (\ref{l}) as
\begin{equation}
\frac{\dot{r}^2}{\dot{\phi}^2}=\left(\frac{dr}{d\phi}\right)^2=\frac{\alpha^2
\sin^2\theta V(r)r^4}{l^2}\left[(\epsilon+a^2)M^2-q^2V(r)\right]
-\alpha^2\sin^2\theta r^2 \ .
\end{equation}
Defining a new variable $u=1/r$ we can express the above equation in a simpler
form:
\begin{equation}
\label{u-eq}
\left(\frac{du}{d\phi}\right)^2=A-Bu-Cu^2 \ ,
\end{equation}
with
\begin{equation}
A=\frac{\alpha^2M^2}{J^2}\left[(\epsilon+a^2)-\frac{q^2}{M^2}\right] \ ,
\end{equation}
\begin{equation}
B=\frac{4m\alpha M^2}{J^2}\left[(\epsilon+a^2)-2\frac{q^2}{M^2}\right] \ ,
\end{equation}
and
\begin{equation}
C=\frac{\omega^2+\alpha^2l^2}{J^2} \ .
\end{equation}
Admitting that the solution of (\ref{u-eq}) has the form:
\begin{equation}
\label{u}
u(\phi)=D+E\cos(\lambda\phi) \ ,
\end{equation}
we found that the constants are given by:
\begin{equation}
D=-\frac{2mM^2\alpha}{\omega^2+\alpha^2l^2}\left[\epsilon+a^2-2\frac{q^2}{M^2}
\right] \ ,
\end{equation}
\begin{equation}
E^2=\frac{4m^2M^4\alpha^2}{(\omega^2+\alpha^2l^2)^2}\left[\epsilon+a^2-2\frac{q^2}
{M^2}\right]^2+\frac{\alpha^2M^2}{\omega^2+\alpha^2l^2}\left[\epsilon+a^2
-\frac{q^2}{M^2}\right] 
\end{equation}
and
\begin{equation}
\lambda^2=\frac{\omega^2+\alpha^2l^2}{J^2} \ ,
\end{equation}
which is smaller than unity.

Finally in order to have trajectories equation unbounded from below, i.e., that 
admit that $r$ goes to infinity, we must have $|E|\geq|D|$, which imply $\epsilon
+a^2-q^2/M^2\geq 0$. So this equation of motion corresponds to the movement of a
test particle constrained to a cone, where its radial coordinate increases without
limit.

\section{Concluding Remarks}

In this work we have presented an exact solutions of five-dimensional Einstein 
equation which admits a magnetic monopole in a point-like global monopole spacetime. 
Our solution is a generalization of the previous ones found by Gross and Perry and
Barnerjee {\it et all}. The latter in the ${\cal{M}}=0$ limit. 

Although the solution presented by Gross and Perry corresponds to a point-like
configuration of magnetic monopole, it is a regular solution in the sense that it 
has a finite inertial mass. Our solution, on the other hand, is valid only in the
region outside the global monopole. Admitting a point-like configuration to the
latter, we can observe that because $g_{00}=-1$, the gravitational mass associated 
with our solution is zero, although it possesses a finite inertial mass.

As we have said our solutions to the components of the metric tensor and the radial
function $V(r)$, were obtained in the region outside the global monopole's core. In 
a pure global monopole system in four-dimensional spacetime, the exact solution for 
the equations of motion, considering the region near the monopole's core, can only 
be obtained numerically \cite{H-L,O}. So we do not expect to find for this 
more general system analytical solutions either. Moreover, as to the global monopole
system, numerical calculation indicates the existence of a small negative 
gravitational mass \cite{H-L} to this object. On the other hand, considering also 
the presence of a Non-Abelian magnetic monopole in the system, positive effective 
gravitational mass to this composite topological object has been found 
\cite{Spi,EYB}. So, these aspects suggest similar properties to this present 
composite monopole. These are points to be investigated in the future. 

In spite of our solution to the metric tensor (\ref{comp}) presents an 
explicit dependence on the parameter $a$, we cannot say that we have found a
family of independent solutions. The numerical factors which appear in those
components can be gauged away by a redefinition of a coordinate system. Consequently
all the invariants of the respective manifold do not depend on it.

By a direct calculation we found that the solid angle associated with the space
section of (\ref{metric1}) presents a radial dependence. This fact
is a consequence of the long range effect of the radial function $V(r)=1+
4m/\alpha r$. The solid angle is:
\begin{equation}
\Omega=4\pi\alpha^2\frac1{\left[1+\frac{2m}{\alpha r}V^{-1/2}(r)\ln(1+\frac{\alpha r}
{2m}\left(1+V^{1/2}(r)\right)\right]^2} \ ,
\end{equation}
which asymptotically reproduces the well known result found in the pure global
monopole spacetime. So at spatial infinity (\ref{metric1}) approaches to the
global monopole metric solution (\ref{gm}) in the presence of the Dirac magnetic 
monopole.

The analysis of the classical trajectories of a massive charged  particles in this
manifold has also been performed. We observe that, as in the flat three-dimensional 
case, this particle has its motion confined to a cone with radial coordinate 
increasing without limit, indicating that there is no bound states.

{\bf{Acknowledgments}}
\\       \\
We would like to thank to C. Galv\~ao and A. C. V. V. de Siqueira for useful 
discussions. Also we would like to thank Conselho Nacional de Desenvolvimento 
Cient\'\i fico e Tecnol\'ogico (CNPq.) for partial financial support.

\end{document}